PAPER • OPEN ACCESS

# New results by low momentum approximation from relativistic quantum mechanics equations and suggestion of experiments

# Journal of Physics Communications

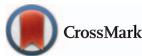

# New results by low momentum approximation from relativistic quantum mechanics equations and suggestion of experiments

Huai-Yu Wang

Department of Physics, Tsinghua University, Beijing 100084, People's Republic of China

**E-mail:** wanghuaiyu@mail.tsinghua.edu.cn

## Abstract

A fundamental belief is that the formulism of relativistic quantum mechanics equations (RQMEs) should remain in low momentum motion. However, it is found that some formulas from RQMEs were lost in Schrödinger equation. For example, a free relativistic particle has positive and negative energy branches. The former includes positive kinetic energy (PKE) and the latter negative kinetic energy (NKE). The latter should be treated on an equal footing as the former. Nevertheless, from Schrödinger equation, a free particle can have only PKE. Starting from RQMEs and taking low momentum approximation, we derive NKE Schrödinger equation which is for the cases that free particles have NKE. Thus negative energy branch of RQMEs can be retained in low momentum motion. We point out a fact that whether Schrödinger equation is applicable in a region where a particle's energy $E$ is less than potential $V$, $E < V$, has never been quantitatively verified. In such a region NKE Schrödinger equation should be employed. With the help of NKE Schrödinger equation, the lost formulas are recovered. The so-called difficulty of negative probability of Klein–Gordon equation for free particles is solved. A PKE (NKE) particle can have stationary motion only when it is subject to an attractive (repulsive) potential, which is determined by Virial theorem. Two NKE electrons in a potential can constitute a stable system, a new kind of possible mechanism for electron paring. The whole discussion stems from RQMEs with no any new postulation. Experiments are suggested, which may confirm that there are indeed NKE electrons.

## 1. Introduction

At the very beginning when Schrödinger established the equation entitled by his name, he tried to extend his equation to the case of relativity [1], but failed, for the evaluated fine structure of hydrogen atom was not consistent with experimental one [2]. Thus the extension from low momentum quantum mechanics (QM) equation to relativistic one was not successful. Later, relativistic quantum mechanics equations (RQMEs) were established. The RQMEs refer to Klein–Gordon equation and Dirac equation which respectively described the motion of particles with spin-0 and 1/2. Starting from either one of the RQMEs, Schrödinger equation could be obtained after low momentum approximation was taken. Therefore, compared to Schrödinger equation, the RQMEs are more fundamental ones.

The author thinks that the failure of the extension from Schrödinger equation to relativistic case hints that something might be lost in low momentum motion. Hence, we like to ask a question: when RQMEs were approximated to Schrödinger equation, the low momentum motion, was there anything lost? The author thinks that the concept of negative kinetic energy (NKE) may be one.

In the author's opinion, all the formalism and features of RQMEs should be retained in their forms of low momentum approximation. However, we do find that some have been lost, as will be seen in section 2 below. The author thinks that this is due to the neglection of the concept of NKE.

It is well known that, according to RQMEs, the expression of a free particle's energy includes positive and negative branches. Each owns corresponding eigen functions. The negative energy branch contains NKE. That is to say, free microscopic particles may have either positive kinetic energy (PKE) or NKE.

In classical mechanics, a free body is of positive kinetic energy. After QM had appeared, people inherited this idea. The negative kinetic energy of a free particle was incompatible with peoples' common sense and one could not tell the corresponding physical image and physical meaning.

So far, physicists have been generally afraid of NKE of free particles, in spite of that RQMEs told us that a free particle had solutions of negative energy. People tried to avoid and ignore the NKE as far as possible. A way to avoid the NKE was to understand Klein–Gordon equation as a field equation the quantized energies of which were always positive even though the parameter $E$ in the wave function could be either positive or negative [3]. However, the negative energy difficulty was not removed in this way, since it was only removed on quantisation [4].

The author thinks that since we acknowledge that the RQMEs have PKE and NKE spectra, the NKE ought to be treated on an equal footing with the PKE. It is believed that the NKE has its own physical significance.

Here we mention that any mathematical detail should be seriously faced to, because it is of corresponding physical implications. Correct mathematics reflects correct physical meanings. For example, Planck proposed his formula describing black-bodies radiation. The calculated curve was perfectly in agreement with experiments, which meant the mathematics was correct. Correspondingly, it had significant physical meaning.

The present work will make the low momentum approximation from the RQMEs, so as to obtain the equation of a particle with NKE in low momentum motion. In this equation, the operator of kinetic energy has a minus sign, call NKE operator. The expectation value of the NKE operator in a state is always nonpositive. Through this procedure, particles' NKE is retained in low momentum motion.

Suppose a particle's energy is $E$. It is able to go through a region with potential $V$ where $V > E$. This is well-known tunneling effect. The particle's energy $E$ is its kinetic energy plus potential $V$. Since $E < V$, its kinetic energy has to be negative. People usually avoid discussing NKE. Some ones might make a far-fetched explanation of the tunnel effect in terms of uncertainty relationship. The uncertainty relationship concerns the degree of inaccuracy of our measurements of a particle's position and momentum. No matter whether we carry out a measurement or not, that the particle does pass through the potential region is a fact. As long as a particle is in such a potential area, its kinetic energy is bounded to be negative.

Let us point out a fact. In QM textbooks, the reflection and transmission coefficients are evaluated by Schrödinger equation when a particle goes through a potential barrier in the case that $E < V$, but this evaluation method has never been quantitatively verified. This means that in an experiment there is a potential barrier and a particle with $E < V$ goes through it. One measures the transmission coefficient. Then he calculates the coefficient theoretically by means of the method as in QM textbooks, and the calculated value is in agreement with the measured one quantitatively. There is no such literature. No QM textbook has claimed that the method of evaluating the transmission coefficient of a particle's tunneling was quantitatively verified by experiments. This fact manifests that whether Schrödinger equation is applicable or not in the region $E < V$ has never been quantitatively verified.

The viewpoint that a particle in a potential barrier is of NKE must have experimental confirmation.

This paper deals with topics within the scope of RQMEs, and does not touch topic of interpretation of QM and quantum gravity.

The layout of this paper is as follows. In section 2, we point out that after low momentum approximation is made from RQMEs, some formulas are lost. Making low momentum approximation in the case of negative energy branch, we obtain NKE Schrödinger equation. In this way, the lost formulas are recovered. In order to remedy the relationship involving probability current, Klein–Gordon equation is decoupled to be PKE and NKE ones, so as to solve the difficulty of negative probability derived from Klein–Gordon equation. Thus, we affirm the existence of NKE in low momentum motion based on the RQMEs. The well-known Schrödinger equation can only describe the motion of particles in regions where they are of PKE. In section 3, it is shown that a NKE particle can have a stable motion only when it is subject to a repulsive potential. The reason is that stable motion should obey Virial theorem. In section 4, four experiments are suggested that may verify the existence of NKE electrons. Section 5 is our concluding remarks. In appendix A, the current densities of PKE and NKE solutions solved from Dirac equations are given, showing that those of PKE and NKE particles are just opposite in direction. In appendix B, it is disclosed that in each RQME, when potential changes its sign, the eigen values do either. In appendix C, the NKE solution of Dirac equation with a repulsive Coulomb potential is emphasized. In appendix D, it is pointed out that the NKE particles are believed dark paticles, and there are related topics to be researched. In this appendix, some, but not all, works in undertaking or to be done are briefly outlined.

## 2. The negative kinetic energy of particles

### 2.1. Three point worthy of mention in Schrödinger equation

Schrödinger equation can be derived from RQMEs. The RQMEs refer to Klein–Gordon equation

$$\left(i\hbar\frac{\partial}{\partial t} - V\right)^2 \psi = (m^2c^4 - c^2\hbar^2\nabla^2)\psi \tag{2.1}$$

and Dirac equation

$$(\gamma^\mu \partial_\mu + V\beta + im)\psi = 0. \tag{2.2}$$

They respectively describe the relativistic motion of particles with spin 0 and 1/2. Through out this work, we assume that the potential $V$ is independent of time $t$.

Now we take the transformation [5, 6]

$$\psi = \psi_{(+)} e^{-imc^2 t/\hbar} \tag{2.3}$$

in equations (2.1) and (2.2) and make low momentum approximation. Then we obtain Schrödinger equation,

$$i\hbar \frac{\partial \psi_{(+)}}{\partial t} = \left(-\frac{\hbar^2}{2m}\nabla^2 + V\right)\psi_{(+)}. \tag{2.4}$$

The corresponding stationary equation is

$$\left(-\frac{\hbar^2}{2m}\nabla^2 + V\right)\psi_{(+)} = E_{(+)}\psi_{(+)}. \tag{2.5}$$

Please note that in equations (2.3) and (2.4) $\psi_{(+)}$ depends on space coordinates and time, but in equation (2.5), $\psi_{(+)}$ depends merely on space coordinates. We do not explicitly write the variables.

Here it should be emphasized that the operator of kinetic energy is $-\frac{\hbar^2}{2m}\nabla^2$. Its expectation in a state is nonnegative. Therefore, this operator is called PKE operator and we say the particle's kinetic energy is positive. Thus, equations (2.4) and (2.5) can also be called PKE Schrödinger equation. A subscript (+) is attached to the wave function and energy showing that they belong to PKE Schrödinger equation.

It is believed that all the properties the relativistic motion has should be retained in low momentum approximation. However, there are at least three points we like to mention in low momentum approximation from equations (2.1) and (2.2) to (2.4) and (2.5)

The first point is the negative energy branch of a free particle.

A free particle obeying RQMEs is of eigen energies

$$E_{(\pm)} = \pm\sqrt{m^2c^4 + p^2c^2}. \tag{2.6}$$

It has two branches. The energies $\pm mc^2$ when momentum is zero are called static energies. In each energy branch, after removal of the static energy, the remaining part is called the kinetic energy of the particle, denoted as $K$. Then the positive and negative kinetic energies are respectively

$$K_{(+)} = \sqrt{m^2c^4 + p^2c^2} - mc^2 > 0 \tag{2.7a}$$

and

$$K_{(-)} = -\sqrt{m^2c^4 + p^2c^2} + mc^2 < 0. \tag{2.7b}$$

When in low momentum, the energy and kinetic energy are approximately

$$E_{(+)} = mc^2 + \frac{p^2}{2m}, \quad K_{(+)} = \frac{p^2}{2m} > 0 \tag{2.8a}$$

and

$$E_{(-)} = -mc^2 - \frac{p^2}{2m}, \quad K_{(-)} = -\frac{p^2}{2m} < 0. \tag{2.8b}$$

Please note that for equations (2.7a) and (2.8a), there is always $E_{(+)} > 0$ and $K_{(+)} > 0$; while for (2.7b) and (2.8b), there is always $E_{(-)} < 0$ and $K_{(-)} < 0$.

However, Schrödinger equation (2.5) with no potential has only energy (2.8a). That is to say, equation (2.8b) is lost in low momentum motion. The negative energy of Dirac equation was interpreted as the solution of an antiparticle. An antiparticle has a positive kinetic energy. However, here we are talking about the energy (2.8b) which is negative for low momentum motion.

Let us take a consideration as follows. The variation of a particle momentum can be continuous. Suppose that there is a relativistic free particle with negative energy, and we can, by means of some way, let its momentum become less, until it does low momentum motion. Then, in low momentum motion, its energy should remain negative. If its negative energy disappears, we have to answer at what momentum the negative energy disappears and why. One cannot tell a reason of the possible disappearance of the negative energy in the course of the continuous reduction of momentum. However, from Schrödinger equation, one is unable to solve a free particle with negative energy.

The second point is the relationship between the probability currents of particles with positive and negative energies.

The eigen wave functions belonging to $E_{(\pm)}$ are denoted as $\psi_{(\pm)}$ and the corresponding current densities denoted as $\boldsymbol{j}_{(\pm)}$. We calculate $\boldsymbol{j}_{(+)}$ and $\boldsymbol{j}_{(-)}$ for Dirac equation, see appendix A. The results show that

$$\boldsymbol{j}_{(-)} = -\boldsymbol{j}_{(+)}. \tag{2.9}$$

The direction of probability current of particles with positive energy $E_{(+)}$ is opposite to that with negative energy $E_{(-)}$.

Let us inspect Schrödinger equation (2.4). The probability current was derived from continuity equation

$$\frac{\partial \rho}{\partial t} + \nabla \cdot \boldsymbol{j} = 0. \tag{2.10}$$

For Schrödinger equation, the probability density was defined by

$$\rho_{(+)} = \psi_{(+)}^* \psi_{(+)}. \tag{2.11}$$

Substitution of (2.4) and (2.11) into (2.10) gave the current density

$$\boldsymbol{j}_{(+)} = -\frac{i\hbar}{2m}(\psi_{(+)}^* \nabla \psi_{(+)} - \psi_{(+)} \nabla \psi_{(+)}^*). \tag{2.12}$$

There is no expression for $\boldsymbol{j}_{(-)}$. So, the formula (2.9) is lost in low momentum motion.

The third point concerns the change of the sign of the potential in RQMEs.

In RQMEs (2.1) and (2.2) for stationary motion, if the potential $V$ changes its sign, then the obtained eigenvalues also change sign. This property is denoted as follows.

$$\text{If } V \to -V, \text{ then } E \to -E. \tag{2.13}$$

The proof is shown in appendix B. Apparently, Schrödinger equation (2.5) does not have the property (2.13).

## 2.2. Negative kinetic energy Schrödinger equation

The author firmly believe that all the properties in relativistic motion ought to be retained when low momentum motion is taken. In fact, we are able to recover properties (2.8b), (2.9) and (2.13) in low momentum motion. The way to do so is as follows.

Formalism in physics should exhibits symmetry. Beside the transformation (2.3), we have another one,

$$\psi = \psi_{(-)} e^{imc^2 t/\hbar}. \tag{2.14}$$

When (2.14) is substituted into equations (2.1) and (2.2) and low momentum approximation is made, we obtain

$$i\hbar \frac{\partial \psi_{(-)}}{\partial t} = \left(\frac{\hbar^2}{2m}\nabla^2 + V\right)\psi_{(-)}. \tag{2.15}$$

The equation for stationary motion is

$$\left(\frac{\hbar^2}{2m}\nabla^2 + V\right)\psi_{(-)} = E_{(-)}\psi_{(-)}. \tag{2.16}$$

Compared to Schrödinger equation, in equations (2.15) and (2.16) the operator of kinetic energy has a minus sign, $\frac{\hbar^2}{2m}\nabla^2$. This means that its expectation in a state is non-positive. This operator is called NKE operator, and equations (2.15) and (2.16) are called NKE Schrödinger equation. In this case, we say that the particle's kinetic energy is negative. A subscript ( − ) is attached to the wave function and energy showing that they belong to NKE Schrödinger equation.

With the help of NKE Schrödinger equation, we are able to recover the above mentioned three points, equations (2.8*b*), (2.9) and (2.13).

First, in equation (2.16), when the potential is absent, the energy is (2.8b).

Second, the probability density of equation (2.16) is defined by

$$\rho_{(-)} = \psi_{(-)}^* \psi_{(-)}. \tag{2.17}$$

Substitution of (2.15) and (2.17) into continuity equation (2.10) gives

$$\boldsymbol{j}_{(-)} = \frac{i\hbar}{2m}(\psi_{(-)}^*\nabla\psi_{(-)} - \psi_{(-)}\nabla\psi_{(-)}^*). \tag{2.18}$$

Comparison of equations (2.12) and (2.18) shows that the property (2.9) is retrieved. This property reveals that the probability current of NKE particles are opposite to that of PKE particles.

Third, in both equations (2.5) and (2.16) if the potential $V$ is replaced by $-V$, one immediately sees that $E_{(-)} \to -E_{(+)}$. Thus, the property (2.13) is recovered.

Thus, both the properties (2.8b), (2.9) and (2.13) are retained in low momentum motion. The prerequisite is that NKE Schrödinger equation is essential in describing motion of particles in QM. This provokes us to think the concept of NKE. Let us make some discussions with respect to equations (2.7) and (2.8).

First, both positive and negative energies are of physically significant existences and the negative one should not be thrown away at will. Equations (2.7a) and (2.8a) demonstrate that the total energy of a free particle is the sum of the positive static and kinetic energies. When moving with low momentum, the constant term $mc^2$ does not affect physical processes so that can be dropped, and the motion is described by Schrödinger equation with the kinetic energy operator in equation (2.4).

Equations (2.7b) and (2.8b) demonstrate that the total negative energy is the negative static energy plus NKE. Up to now, the negative energy has not been paid enough attention. In classical mechanics, kinetic energy is indeed impossible to be negative. However, equation (2.15) is from RQMEs. The negative energy branch should be retained in low momentum approximaition. Some quantities in QM may not have their classical correspondences.

The operators in Schrödinger equation have their classical analog: $i\hbar\frac{\partial}{\partial t}$ and $-\frac{\hbar^2}{2m}\nabla^2$ respectively representing total energy and kinetic energy. However, in equations (2.15) and (2.16) the operator $\frac{\hbar^2}{2m}\nabla^2$ does not have classical correspondence. What behavior will be when a PKE particle interacts with a NKE one needs further study.

Second, in Newtonian mechanics, kinetic energy is defined as the square of the momentum divided by twice the mass, $p^2/2m$. However, from the theory of relativistic QM, the kinetic energy is defined by equation (2.7). Equation (2.8) should not be thought as the basic definition of kinetic energy, since they are just approximate forms in low momentum motion. Therefore, when speaking of NKE (2.8b), one should realize that

$$p^2/2m \neq (i\mu)^2/2m. \tag{2.19}$$

There is no way to derive an equation like $p^2/2m = (i\mu)^2/2m$ from equations (2.7) and (2.8) for a particle's energy less than potential $E < V$.

When there is a potential, the energies in equation (2.7) become

$$E_{(+)} = \sqrt{m^2c^4 + p^2c^2} + V > V \tag{2.20a}$$

and

$$E_{(-)} = -\sqrt{m^2c^4 + p^2c^2} + V < V. \tag{2.20b}$$

Their low momentum approximations are

$$E_{(+)} = mc^2 + K_{(+)} + V > V \tag{2.21a}$$

and

$$E_{(-)} = -mc^2 + K_{(-)} + V < V. \tag{2.21b}$$

Please note that for equations (2.20a) and (2.21a), there is always $E_{(+)} > V$; while for (2.20b) and (2.21b), there is always $E_{(-)} < V$.

Both the positive and negative energy branches have low momentum approximations as shown by equations (2.20) to (2.21). For the case of PKE (2.21a), i. e., $E_{(+)} > V$, the transformation for low momentum approximation should be equation (2.3); while in the case of NKE (2.21b), i. e., $E_{(-)} < V$, the transformation should be (2.14).

Since Schrödinger equation (2.5) was established, the expression of $K = p^2/2m$ for kinetic energy has been extended to the case of $K < 0$, although no one has rigorously proved it. Here we stress that the expression of the kinetic energy for $K < 0$ should be derived from equations (2.7b) and (2.8b). This means that a particle doing low momentum does not have imaginary momentum, as shown by equation (2.19), even its energy $E$ is less than potential $V$.

The basic definition of positive kinetic energy comes from relativistic mechanics, and the basic concept of NKE comes from relativistic QM.

There was a work [7] which thought that in classically forbidden regions the NKE of a particle might be measured. The measured NKE was caused by measurement error. That analysis discussed measurement theory with regard to Schrödinger equation. Here we are talking about the equation itself and there is a NKE operator in the equation, not concerning measurement theory.

### 2.3. NKE Decoupled Klein–Gordon equation

Equation (2.4) applies to the cases of a particle having PKE and equation (2.15) to that having NKE. Their corresponding probability currents are equations (2.12) and (2.18) respectively.

Equation (2.9) is valid both for Dirac equation and low momentum motion. Then what about Klein–Gordon equation? This question is closely related to the so-called difficulty of 'negative probability' of Klein–Gordon equation. In this subsection, we merely consider the cases where potential is piecewise constant.

The difficulty of 'negative probability' [8, 9] emerged soon after Klein–Gordon equation had been proposed. The appearance of this difficulty was closely related to the definition of the probability current. When equation (2.1) and probability current (2.12) were substituted into (2.10), one found that expression of probability density for Klein–Gordon equation had to be

$$\rho = \frac{i\hbar}{2mc^2}\left(\Psi^*\frac{\partial}{\partial t}\Psi - \Psi\frac{\partial}{\partial t}\Psi^*\right). \tag{2.22}$$

Since the first derivative of the wave function with respect to time was involved in equation (2.22), one was unable to guarantee that the probability was definitely positive, which was so-called difficulty of negative probability. Furthermore, the density would be discontinuous at the place where potential had a jump [10]. This difficulty arose from the second time derivative in Klein–Gordon equation.

As a matter of fact, in the cases of Dirac equation and low momentum motion, the probability densities (2.11) and (2.17) were defined first, and then the expressions of $j_{(+)}$ and $j_{(-)}$ were derived. While for Klein–Gordon equation, the probability current (2.12) was firstly assumed from which the probability density equation (2.22) was derived. The author thinks that this procedure was illegal.

Since the continuity equation (2.10) contains the first time derivative of density, it should be drawn from an equation that contains merely the first time derivative of wave function.

Although Klein–Gordon equation (2.1) does contain quadratic derivative to time, it is possible to recast it into the form containing first time derivative. This can be done as follows.

When the potential is a piecewise constant one, Klein–Gordon equation equation (2.1) can be rewritten as

$$\left(i\hbar\frac{\partial}{\partial t} - H_{(-)}\right)\left(i\hbar\frac{\partial}{\partial t} - H_{(+)}\right)\psi = 0, \tag{2.23}$$

where we have denoted that

$$H_{(\pm)} = \pm H_0 + V, \quad H_0 = \sqrt{m^2c^4 + c^2(-i\hbar\nabla)^2}. \tag{2.24}$$

Now, let us impose a restriction that

$$\left(i\hbar\frac{\partial}{\partial t} - H_{(+)}\right)\psi_{(+)} = 0. \tag{2.25}$$

This equation actually has been presented in textbools [11–13]. It was called Salpeter equation and studied with some specific potentials [14–28]. This form is helpful for clarifying the concept of probability current.

The probability density is defined by equation (2.11). It is easy to put down

$$\frac{\partial \rho_{(+)}}{\partial t} = \frac{1}{i\hbar}(\psi_{(+)}^* H_{(+)}\psi_{(+)} - \psi_{(+)} H_{(+)}\psi_{(+)}^*). \tag{2.26}$$

We take Taylor expansion of $H_0$ in equation (2.24),

$$\sqrt{m^2c^4 - c^2\hbar^2\nabla^2} = mc^2 + mc^2\sum_{n=1}^{\infty} b_n \nabla^{2n}, \tag{2.27a}$$

where

$$b_1 = -\frac{\hbar^2}{2m^2c^2}, \quad b_n = \frac{(2n-3)!!}{2^n n!}\left(\frac{\hbar^2}{m^2c^2}\right)^n, \quad n \geqslant 2. \tag{2.27b}$$

The probability current of the $n$-th order that depends on wave function is defined as

$$\mathbf{j}_{(1)}(\psi) = -\frac{i\hbar}{2m}(\psi^*\nabla\psi - \psi\nabla\psi^*) \tag{2.28a}$$

and

$$\boldsymbol{j}_{(n)}(\psi) = \frac{i\hbar}{2m} b_n \sum_{i=0}^{n} (-1)^i [\nabla^i \psi^* \nabla^{2n-i-1} \psi - \nabla^i \psi \nabla^{2n-i-1} \psi^*]. \tag{2.28b}$$

Then, the expression of probability current is

$$\boldsymbol{j}(\psi) = \boldsymbol{j}_{(1)}(\psi) + \boldsymbol{j}_{(2)}(\psi) + \cdots = \sum_{n=1}^{\infty} \boldsymbol{j}_{(n)}(\psi). \tag{2.28c}$$

If we denote

$$\boldsymbol{j}_{(+)} = \boldsymbol{j}(\psi_{(+)}), \tag{2.29}$$

then, probability current (2.28) and probability density (2.11) meet the continuity equation (2.10),

On the other hand, we may also exchange the order of the two factors in (2.23), and impose a restriction that

$$(i\hbar \frac{\partial}{\partial t} - H_{(-)}) \psi_{(-)} = 0. \tag{2.30}$$

The particles' probability density is defined by equation (2.17). Then one immediately finds that the probability current

$$\boldsymbol{j}_{(-)} = -\boldsymbol{j}(\psi_{(-)}). \tag{2.31}$$

meets equation (2.10).

Here we mention that equation (2.28) has in fact been obtained before by means of (2.11), (2.26) and (2.27), see equation (3.23) in [28]. However, here we also give the expression of $\boldsymbol{j}_{(-)}$ for NKE branch, equation (2.31). Equations (2.29) and (2.31)show than the property (2.9) is met for Klein–Gordon equation.

Equations (2.25) and (2.30)are hereafter respectively called PKE and NKE decoupled Klein–Gordon equations. Under low momentum, they are approximated to be Eq, (2.4) for a particle with PKE and (2.15) for NKE, respectively, and correspondingly, equations (2.28) and (2.31) respectively degrade to (2.12) and (2.18).

From above process, it is seen that at least under a piecewise constant potential the so-called difficulty of negative probability is a false one. We stress that the derivation procedure of obtaining equation (2.22) was inherently wrong. For relativistic motion, the expression of the probability current should be (2.28) instead of (2.12). Misuse of (2.12) lead to the negative probability.

Please note that in equations (2.28) and (2.31), the probability currents are composed of all order terms This is because kinetic energies are so. The terms were obtained by expansion of equation (2.7a). Each term in the kinetic energy expansion has its correspondence in probability current. That only the first term remained, e. g., equation (2.12), lead to the wrong (2.22). This also reminds us that when we write down kinetic energy, if only the first term is retained, i. e., (2.28a), problems will probably be yielded. Thus, we know that for Klein-Gorden equation, the expression of the probability current is not the simple form of equation (2.12) as believed by researchers [29, 30], but has a complex forms.

By the way, it was shown that decoupled Klein–Gordon equation (2.25) was of relativistic invariance [11].

In short, a rule is that the expression of probability density is always the transpose conjugate of wave function multiplied by itself, $\rho = \psi^\dagger \psi$. Then, the probability current is determined by the continuity equation. The expression of the probability current depends on Hamiltonian. For a relativistic system with spin of 1/2, the probability current is defined as (A.4), (A.10) or (A.22) in one-, two- or three- dimensional space, where the currents of both PKE and NKE particles are automatically included; for a relativistic system with spin zero, the currents of PKE and NKE particles are respectively (2.28) and (2.31); for a system with spin zero doing low momentum motion, the currents of PKE and NKE particles are (2.12) and (2.18), respectively.

## 3. The stable solutions of NKE Schrödinger equation

It was seen in the last section that in QM, there was a property (2.13). Here we give two examples.

The first example is infinitely high potential barrier.

The potential in equation (2.16) is

$$V(x) = \begin{array}{ll} 0, & 0 \leqslant x \leqslant a \\ -\infty & x < 0, x > a. \end{array} \tag{3.1}$$

This is one-dimensional infinitely high barrier. Obviously, eigen functions of equation (2.16) with the potential (3.1) are exactly the same as those of Schrödinger equation (2.5) with potential $-V(x)$, and the eigenvalues are contrary numbers to each other. So we immediately put down the eigen wave functions

$$\psi_{(-)} = \sqrt{\frac{2}{a}} \sin \frac{n\pi}{a} \qquad (3.2a)$$

and eigenvalues

$$E_{(-)n} = -\frac{n^2 \hbar^2 \pi^2}{2ma^2}, \; n = 1, 2, \cdots. \qquad (3.2b)$$

When a particle with NKE is in the infinitely high potential barrier, the highest energy level in equation (3.2) is closest to the top of the barrier. The energy spectrum has an upper limit but no lower limit of the NKE [31].

A PKE particle is bounded in the well, while a NKE one is bounded in the barrier.

The second example is Coulomb potential. If an electron is subject to a Coulomb attractive potential, Schrödinger equation equation is

$$\left( -\frac{\hbar^2}{2m}\nabla^2 - \frac{e^2}{4\pi\varepsilon_0 r} \right) \psi_{(+)} = E_{(+)} \psi_{(+)}. \qquad (3.3)$$

Its eigen energies for bounded states are

$$E_{(+)n} = -\frac{\varepsilon}{n^2}, \; \varepsilon = \frac{me^4}{2\hbar^2}, \; n = 1, 2, \cdots. \qquad (3.4)$$

The discrete spectrum has a lowest limit $E_{(+)1} = -\varepsilon$ and no upper limit.

If in NKE Schrödinger equation (2.16) the Coulomb potential is a repulsive one,

$$\left( \frac{\hbar^2}{2m}\nabla^2 + \frac{e^2}{4\pi\varepsilon_0 r} \right) \psi_{(-)} = E_{(-)} \psi_{(-)}. \qquad (3.5)$$

Its discrete eigen energies are

$$E_{(-)n} = \frac{\varepsilon}{n^2}, \; \varepsilon = \frac{me^4}{2\hbar^2}, \; n = 1, 2, \cdots. \qquad (3.6)$$

They are also bounded states. A NKE particle can have stationary motion when subject to a repulsive potential.

The solutions of Dirac equation with a Coulomb potential are presented in appendix C. The low momentum approximation of the spectrum (C.10) is (3.4), which has a lower limit but no upper limit. That of the spectrum (C.14) is (3.6), which has an upper limit but no lower limit. Hence, both the positive and negative energy branches have their low momentum approximations.

The above two examples reveal two features. One feature is that when a particle is inside a potential barrier, it is of NKE, which is actually manifestation of (2.20b) and (2.21b). The other feature is that a NKE particle can have bounded states only when subject to a repulsive states. Please note that the discrete spectra (3.2b) and (3.6) mean bounded states. This feature seems contradictory to our common sense. Intuitively, a particle can have stable states only when it is attracted by others. However, this common sense is based on that the particle's kinetic energy is positive.

Whether a system is able to reach a stable state or not can be recognized by the following thinking. A system's energy is the sum of two parts: kinetic energy and potential. When one part is positive and the other is negative, due to the competition between the two parts, it is possible for the system's total energy to reach an equilibrium point, so that the system has stable motion. If both parts are positive or negative, there is no possibility to reach such an equilibrium point. This is clearer by Virial theorem.

Virial theorem is derived from

$$\frac{d\langle \boldsymbol{r} \cdot \boldsymbol{p} \rangle}{dt} = \frac{1}{i\hbar}\langle [\boldsymbol{r} \cdot \boldsymbol{p}, H] \rangle = 2\langle T \rangle - \langle \boldsymbol{r} \cdot \nabla V \rangle = 0, \qquad (3.7)$$

where it can be evaluated that $[\boldsymbol{r} \cdot \boldsymbol{p}, T] = [\boldsymbol{r} \cdot \boldsymbol{p}, \frac{1}{2m}(p_x^2 + p_y^2 + p_z^2)] = 2i\hbar T$ and $[\boldsymbol{r} \cdot \boldsymbol{p}, V] = -i\hbar \boldsymbol{r} \cdot \nabla V$. Therefore, Virial theorem is

$$2\langle T \rangle = \langle \boldsymbol{r} \cdot \nabla V \rangle. \qquad (3.8)$$

We take Coulomb potential as an example, i. e., $V = a/r$. Then Virial theorem becomes

$$2\langle T \rangle + \langle V \rangle = 0. \qquad (3.9)$$

If kinetic energy is positive, the potential must be negative, so as to make the sum be zero.

For a NKE system, equation (3.7) becomes

$$2\langle -T \rangle = \langle \boldsymbol{r} \cdot \nabla V \rangle. \qquad (3.10)$$

In the case of Coulomb potential, we have

$$2\langle -T \rangle + \langle V \rangle = 0. \tag{3.11}$$

Since now kinetic energy is negative, only a repulsive potential can meet equation (3.11).

For Dirac equation, the negative energy branch (C.14) has its low momentum approximation (3.6). This indicates that the negative energy branch (C.14) is actually a NKE particle subject to a repulsive Coulomb potential. In appendix B, we point out that in Dirac equation, either (B.5a) or (B.5b) can be employed. Equation (B.5a) gives PKE solutions and (B.5b) gives NKE solutions. The simplest case is that there is no potential, i. e., a free particle. Then the PKE and NKE spectra are (2.6).

That a NKE particle subject to a repulsive potential opens a field of new stable systems where NKE particles participate in. For instance, a PKE electron (proton) and a NKE proton (electron) can constitute a stable system. Such systems will be investigated in our next paper [32].

It is well known that there has been a mechanism of electron pairing disclosed by Cooper, called Cooper pair [33]. In a metal, electrons at Fermi energy level were approximately free, and they collided with lattices. With the help of phonons as a medium, a relatively weak, but net attraction between the two electrons was yielded. This caused the electrons to pair, and the paired two electrons then moved together. The paired electrons are of PKE. That two electrons are pared in a material is the key of the famous BCS superconducting theory [34]. In this kind of superconductors, Cooper pairs were superconducting carriers.

Based on the results in present work, it is possible that there is another mechanism for electron paring.

In solid materials, the arrangement of positive ions constitutes potential wells and walls. In the vicinity of the ions are the wells, and between them are walls. Electrons have probability in the walls where they may be of NKE. Two NKE electrons constitute a pair by means of the Coulomb repulsion between them, which is a stable system. This is a probable pairing mechanics. The paired electrons probably move mainly within potential wells. Since near the ions are wells and between the ions are walls, potential walls connect each other throughout the whole solid material, while the wells may not be so. Hence, the walls, if their thicknesses are appropriate, are likely a path for such kind of electron pairs.

This possible mechanism is of at least two advantages. One is that the higher the potential barriers, the more helpful for the superconductivity of the material. The other is that this mechanics needs not electron-phonon interaction, such that it may be responsible for the superconductivity appearing in materials lacking of phonons such as quasicrystals [35].

Recently, a research work revealed that a new kind of mechanism of electron pairing was possible [36]. In a YBCO thin films, holes with diameters about 100nm were drilled by ion etching method. It was found that electron charges moving around the holes were 2e, indicating that electrons combined into pairs. It was not clear how the electrons formed pairs. Based on the present work, we suggest a possibility as follows. Within some depth of each hole wall, the potential was sufficiently high so that greater than the electrons' energies. Therefore, in such regions, electrons were of NKE, and they naturally constituted pairs. Whether there were really NKE electrons in this film or not can be probably tested by an experiment that will be proposed in the next section.

## 4. Suggestion of experiments

The assertion that there exist NKE particles should be confirmed by experiments. In this section, we suggest four experiments. The fist three experiments concern photon scattering by NKE electrons, which can be done in labs, and the fourth one is to observe celestial spectra.

Before proposing experiments, let us see the effect of the scattering of a photon by a NKE electron.

It is well known that when a photon with wave length $\lambda$ collides with a free rest electron with mass $m$, the wave length $\lambda'$ of the photon at scattering angle $\theta$ is determined by

$$\lambda' = \lambda + \lambda_C(1 - \cos\theta), \tag{4.1}$$

where $\lambda_C = h/mc$ is Compton wave length. It is obvious that some energy of the photon is transferred to the electron during the collision, so that the wave length becomes longer. This is famous Compton scattering [37, 38].

For a free NKE electron, its static energy is negative, $-mc^2$, see equations (2.7b) and (2.8b). Its momentum is negative, which can be understood from two aspects. One is that we have proved [32] that for a NKE body, the velocity $\boldsymbol{v}_{(-)}$ and momentum $\boldsymbol{p}_{(-)}$ have opposite directions, $\boldsymbol{v}_{(-)} = -\boldsymbol{p}_{(-)}/m$. The velocity plays a role to determine the position of the NKE body, while its momentum plays a role to yield physical effect, such as pressure. The other aspect is that the probability current of NKE particles is opposite to PKE particles, as has been discussed in section 2, e. g., equation (2.9).

Therefore, when a photon is scattered by a NKE electron, the energy conservation is

$$\hbar\omega - mc^2 = \hbar\omega' - \frac{mc^2}{\sqrt{1 - v^2/c^2}}, \tag{4.2}$$

and momentum conservation is

$$\hbar k = \hbar k' \cos\theta - \frac{mv'}{\sqrt{1 - v^2/c^2}} \cos\theta'. \tag{4.3}$$

As a result, we obtain

$$\lambda' = \lambda - \lambda_C(1 - \cos\theta). \tag{4.4}$$

That is to say, the photon acquires energy from the NKE electron during the collision.

Equation (4.4) is to be verified by experiments.

To do such experiments, one probably has to resort to synchrotron radiation which is a device generating photons. Usually, its work windows output photons with known wave lengths, and apparatuses are provided which can measure the wave lengths of scattered photons. In the following first three proposed experiments, the photons of x-ray are assumed.

The first suggested experiment is to let x-ray photons be scattered by tunneling electrons in a scanning tunneling microscopy (STM).

In a STM, the gap between the tip and sample surface is of width of several nanometers. When the electric current is on, the electrons in the gap are tunneling ones. That is to say, they are in a potential barrier, so that they are of NKE inside the gap region. The scenario of experiment is as follows.

A simple and portable STM is made. It can be as simple as possible, because we merely use the tunneling electrons, not to detect the sample surface.

The STM is carried to a work window of synchrotron radiation. When the electric current is on, let the photons of a known wave length $\lambda$ beam the tunneling electrons in the gap. Then, the wave lengths $\lambda'$ of scattered photons are measured. There will be some scattered photons with the wave lengths $\lambda'$ less than $\lambda$, $\lambda' < \lambda$.

To eliminate the influence of the STM instrument other than the tunneling electrons, the experiment can be done when the electric current is off. In this case, there is no tunneling electrons. Thus, the scattering light spectrum when the current is on subtracts that with no current, we have the spectrum yielded by tunneling electrons.

The second suggested experiment concerns the electron pairs in a YBCO film [36].

We mentioned in the last section that in the experiment [36], the electrons within the hole walls were of NKE. Therefore, when colliding with these electrons, photons will acquire energy.

The scenario of experiment is as follows.

The YBCO film sample [36] made is carried to a work window of synchrotron radiation. Let the photons of a known wave length $\lambda$ beam the sample. Then, the wave lengths $\lambda'$ of scattered photons are measured. There will be some scattered photons with the wave lengths $\lambda'$ less than $\lambda$, $\lambda' < \lambda$.

When an external magnetic field is applied, the paired electrons move around the holes. The extra magnetic field may cause some inconvenience of the experiment. In fact, as long as electrons move near the holes, they are of NKE, even if there is no magnetic field. Consequently, without the magnetic field, NKE electrons can still be detected by photon scattering.

In the experiment [36], the YBCO thin film was grown on a SrTO substrate, and then covered by an AAO layer. To eliminate the influence of the substrate and cover layer, the experiment can be done to a sample comprising the SrTO substrate and AAO cover layer, without the YBCO thin film in between. Thus, the scattering light spectrum with the YBCO film in between subtracts that without YBCO film, we have the spectrum yielded by pure YBCO film. Anyhow, as long as the wave lengths of the scattered photons decrease, there are NKE electrons in the YBCO film.

The third suggested experiment is a simplified version of the first one above.

A metal conductor is made a gap by etching, see figure 1. The gap width can be 10 nm or so. When electric current is on, there will be tunneling electrons through the gap. This instrument is carried to a work window of synchrotron radiation to do the photon scattering experiment: let photons beam the gap. The scattering light spectrum with the current on subtracts that with the current off, we have the spectrum yielded by the tunneling electrons.

Compared to a STM, the instrument in this experiment is easier to make, but the gap width is fixed and cannot be tuned.

The fourth suggest experiment is to seek specific celestial light spectra. We have concluded that a PKE electron and a NKE proton could constitute a stable system, called combo hydrogen atom [32]. Its light spectrum can be detected by us. The wave numbers of the spectrum is expressed by

$$\tilde{\nu} = \frac{1 + m/M}{1 - m/M}\tilde{\nu}_0, \qquad (4.5)$$

where $m$ and $M$ are masses of the electron and proton, respectively, and $\tilde{\nu}_0$ is the wave number of hydrogen atom's spectrum on the Earth. Equation (4.5) shows that the spectral lines have blue-shifts compared to those of a hydrogen atom. The author suggests to seek these spectral lines from celestial bodies.

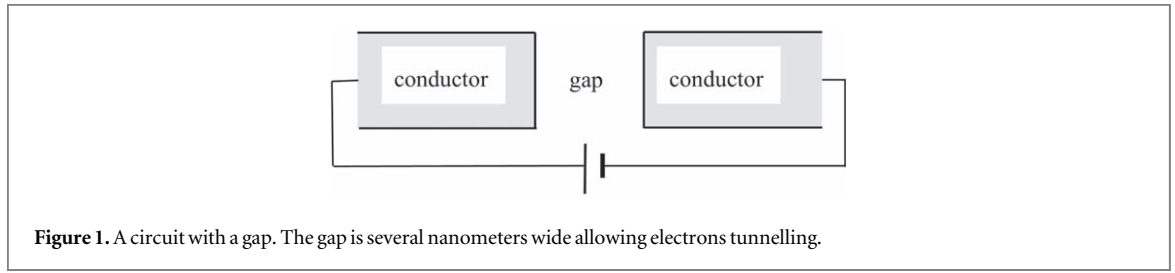

Figure 1. A circuit with a gap. The gap is several nanometers wide allowing electrons tunnelling.

## 5. Concluding remarks

When RQMEs are approximated to Schrödinger equation which describe low momentum motion, some formulas were lost. We list three cases. The first one is that a free relativistic particle has either positive or negative energy, but there is no negative one in Schrödinger equation. The second one is that in relativistic motion the probability currents of PKE and NKE particles are opposite to each other, but there is no such a relationship for Schrödinger equation. The third one is that in RQMEs, when potential takes a minus sign, the eigen energies do either, but this feature was absent in Schrödinger equation.

These lost formulas can be recovered with the help of NKE Schrödinger equation, which is another low momentum approximation from both Dirac equation and Klein–Gordon equation.

The concept of kinetic energy that was from relativistic energy-momentum relationship is reconsidered deliberately. For a relativistic particle, its positive (negative) energy branch contains positive (negative) static energy and kinetic energy. When a particle's energy is less than the potential it is in, it is necessarily of NKE.

Klein–Gordon equation with a piecewise constant potential is decoupled into PKE and NKE ones, from which the relationship between the probability currents of PKE and NKE particles is remedied. The so-called difficulty of negative probability of Klein–Gordon equation is figured out by correctly defining probability current.

Whether a system is stable or not is determined by Virial theorem. A (PKE) NKE particle can have stationary motion only when it is subject to an attractive (repulsive) potential.

Two NKE electrons in a potential can constitute a stable system. This is a new kind of possible mechanism for electron pairing.

The whole discussion stems from RQMEs with no any new postulation.

Four experiments are suggested, among which three can be done in labs. These suggested experiments may verify that there are indeed NKE electrons.

The concept of kinetic energy is one of the earliest in physics. It appeared at the very beginning of classical mechanics, which might solidify people's recognition of this concept. The advent of relativistic quantum mechanics made people realize that the energy of free particles could be negative. So far, however, the NKE has not attracted sufficient attention. People are depressed about the idea that kinetic energy could only be positive, since there has been no reason why the kinetic energy can be negative.

The present paper, based on RQMEs, explores NKE. Our belief is that the NKE and PKE should be treated on an equal footing. From equations (2.6)–(2.8), one hardly tells why one energy branch is more important than the other. Thus, almost all the topics concerning PKE particles should also be re-visited in the aspect of NKE. This inevitably involves a wide range of physical contents. As a matter of fact, the topics of the NKE systems are the contents of our theory of dark matter, which is different from the prevailing dark matter theories. In appendix D, the dark matter theories used nowadays are briefly reviewed, and then we shortly outline the work to be done about the author's NKE theory of dark matter.

## Acknowledgments

This research is supported by the National Key Research and Development Program of China [Grant No. 2016YFB0700102].

## Appendix A. Probability currents of free particles solved from Dirac equations

In this appendix, we show that the currents of NKE wave functions are just opposite to those of PKE ones in terms of one-, two and three-dimensional Dirac equations. The fundamental QM equation is

$$i\hbar \frac{\partial}{\partial t}\psi = H\psi. \tag{A.1}$$

The Hamiltonian of Dirac equation depends on space dimension.

### A.1. One-dimensional space
The Hamiltonian of a free particle is

$$H = -i\hbar c\sigma_1 \frac{d}{dx} + mc^2\sigma_3, \tag{A.2}$$

where

$$\sigma_1 = \begin{pmatrix} 0 & 1 \\ 1 & 0 \end{pmatrix}, \sigma_2 = \begin{pmatrix} 0 & -i \\ i & 0 \end{pmatrix}, \sigma_3 = \begin{pmatrix} 1 & 0 \\ 0 & -1 \end{pmatrix}. \tag{A.3}$$

Equation (A.2) is substituted into (A.1) to achieve continuity equation where the expression of probability current is

$$j = c\psi^+\sigma_1\psi. \tag{A.4}$$

Also from equations (A.2) and (A.1), we get two solutions.

$$E = E_{(+)}, \ \psi_{(+)} = C\begin{pmatrix} pc \\ E_{(+)} - mc^2 \end{pmatrix} e^{i(px - E_{(+)}t)/\hbar}. \tag{A.5}$$

$$E = E_{(-)}, \ \psi_{(-)} = C\begin{pmatrix} E_{(-)} + mc^2 \\ pc \end{pmatrix} e^{i(px - E_{(-)}t)/\hbar}. \tag{A.6}$$

$C$ is a normalization coefficient. By equations (A.4)–(A.6) the probability currents of the two solutions are easily calculated.

$$j_{(+)} = C^2 c\psi_{(+)}^+ \sigma_1 \psi_{(+)} = \frac{pc^2}{E_{(+)}}. \tag{A.7}$$

$$j_{(-)} = C^2 c\psi_{(-)}^+ \sigma_1 \psi_{(-)} = \frac{pc^2}{E_{(-)}} = -j_{(+)}. \tag{A.8}$$

### A.2. Two-dimensional space
The Hamiltonian of a free particle is [39]

$$H = c\boldsymbol{\sigma}\cdot\boldsymbol{p} + mc^2\sigma_3 = \begin{pmatrix} mc^2 & \hbar c\nabla^* \\ -\hbar c\nabla & -mc^2 \end{pmatrix}, \tag{A.9}$$

where $\boldsymbol{\sigma} = \sigma_1 \boldsymbol{e}_x + \sigma_2 \boldsymbol{e}_y$ and $\nabla = i\partial/\partial x + \partial/\partial y$. From equations (A.9) and (A.1), the continuity equation is achieved, where the probability current is expressed by

$$\boldsymbol{j} = c\psi^+\boldsymbol{\sigma}\psi. \tag{A.10}$$

Also from equations (A.9) and (A.1), we get two solutions.

$$E = E_{(+)}, \ \psi_{(+)} = C\begin{pmatrix} (p_x + ip_y)c \\ E_{(+)} - mc^2 \end{pmatrix} e^{i(\boldsymbol{p}\cdot\boldsymbol{r} - E_{(+)}t)/\hbar}. \tag{A.11}$$

$$E = E_{(-)}, \ \psi_{(-)} = C\begin{pmatrix} E_{(-)} + mc^2 \\ (p_x - ip_y)c \end{pmatrix} e^{i(\boldsymbol{p}\cdot\boldsymbol{r} - E_{(-)}t)/\hbar}. \tag{A.12}$$

$C$ is a normalization coefficient. By equations (A.10)–(A.12) the probability currents of the two solutions are easily calculated.

$$\boldsymbol{j}_{(+)} = 2C^2(E_{(+)} - mc^2)(p_x\boldsymbol{e}_x + p_y\boldsymbol{e}_y). \tag{A.13}$$

$$\boldsymbol{j}_{(-)} = 2C^2(E_{(-)} + mc^2)(p_x\boldsymbol{e}_x + p_y\boldsymbol{e}_y) = -\boldsymbol{j}_{(+)}. \tag{A.14}$$

### A.3. Three-dimensional space

The Hamiltonian of a free particle is

$$H = c\boldsymbol{\alpha} \cdot \boldsymbol{p} + mc^2\beta. \tag{A.15}$$

For each energy, there are two solutions respectively with spin up and down.

$$E = E_{(+)}, \psi_{(+)\uparrow} = C(1 \quad 0 \quad b_z \quad b_-)^{\mathrm{T}} e^{i(\boldsymbol{p}\cdot\boldsymbol{r} - |E|t)/\hbar}. \tag{A.16}$$

$$E = E_{(+)}, \psi_{(+)\downarrow} = C(0 \quad 1 \quad b_+ \quad -b_z)^{\mathrm{T}} e^{i(\boldsymbol{p}\cdot\boldsymbol{r} - |E|t)/\hbar}. \tag{A.17}$$

$$E = E_{(-)}, \psi_{(-)\uparrow} = C(-b_z \quad -b_- \quad 1 \quad 0)^{\mathrm{T}} e^{i(\boldsymbol{p}\cdot\boldsymbol{r} + |E|t)/\hbar}. \tag{A.18}$$

$$E = E_{(-)}, \psi_{(-)\downarrow} = C(-b_+ \quad b_z \quad 0 \quad 1)^{\mathrm{T}} e^{i(\boldsymbol{p}\cdot\boldsymbol{r} + |E|t)/\hbar}. \tag{A.19}$$

Here,

$$b = \frac{cp}{|E| + mc^2}; \quad b_\alpha = \frac{cp_\alpha}{|E| + mc^2}, \alpha = x, y, z; \quad b_\pm = b_x \pm ib_y. \tag{A.20}$$

$$C = (1 + b^2)^{-1/2}. \tag{A.21}$$

From equations (A.15) and (A.1), the continuity equation is obtained, where the probability current is expressed by

$$\boldsymbol{j} = c\psi^+ \boldsymbol{\alpha} \psi. \tag{A.22}$$

Equations (A.16)–(A.19) are substituted into (A.22) to evaluate the probability current.

$$\boldsymbol{j}_{(+)\uparrow} = c\psi^+_{(+)\uparrow} \boldsymbol{\alpha} \psi_{(+)\uparrow} = 2C^2 b_z(b_x \boldsymbol{e}_x + b_y \boldsymbol{e}_y + b_z \boldsymbol{e}_z). \tag{A.23}$$

$$\boldsymbol{j}_{(+)\downarrow} = c\psi^+_{(+)\downarrow} \boldsymbol{\alpha} \psi_{(+)\downarrow} = -\boldsymbol{j}_{(+)\uparrow}. \tag{A.24}$$

$$\boldsymbol{j}_{(-)\uparrow} = c\psi^+_{(-)\uparrow} \boldsymbol{\alpha} \psi_{(-)\uparrow} = -\boldsymbol{j}_{(+)\uparrow}. \tag{A.25}$$

$$\boldsymbol{j}_{(-)\downarrow} = c\psi^+_{(-)\downarrow} \boldsymbol{\alpha} \psi_{(-)\downarrow} = -\boldsymbol{j}_{(+)\downarrow}. \tag{A.26}$$

It is seen that in each case, equation (2.9) is satisfied.

## Appendix B. The potential in a relativistic quantum mechanics equation changes its sign

In this appendix, we show that if the potential in a RQME changes its sign, $V \to -V$, then the eigen values also do, $E \to -E$, i. e., equation (2.13).

Klein–Gordon equation for stationary motion is

$$(E - V)^2 \psi = (m^2 c^4 - c^2 \hbar^2 \nabla^2)\psi. \tag{B.1}$$

When the potential is a contrary one, the equation becomes

$$(E + V)^2 \psi = (m^2 c^4 - c^2 \hbar^2 \nabla^2)\psi. \tag{B.2}$$

This is easily rewritten as

$$(-E - V)^2 \psi = (m^2 c^4 - c^2 \hbar^2 \nabla^2)\psi. \tag{B.3}$$

Comparison of equations (B.3) and (B.1) makes us arrive at the conclusion that when the potential becomes a contrary one, the eigenvalues also become contrary ones.

Dirac equation for stationary motion is

$$(c\boldsymbol{\alpha} \cdot \boldsymbol{p} + mc^2\beta + V)\Psi = E\Psi, \tag{B.4}$$

where

$$\boldsymbol{\alpha} = \begin{pmatrix} 0 & \boldsymbol{\sigma} \\ \boldsymbol{\sigma} & 0 \end{pmatrix}. \tag{B.5a}$$

By the derivation procedure of Dirac equation, we are aware of that the $\boldsymbol{\alpha}$ can also be taken as

$$\boldsymbol{\alpha} = \begin{pmatrix} 0 & -\boldsymbol{\sigma} \\ -\boldsymbol{\sigma} & 0 \end{pmatrix}. \tag{B.5b}$$

Now the wave function is written as a two-component spinor, $\Psi = \begin{pmatrix} \varphi \\ \chi \end{pmatrix}$. The two components satisfy the following equations:

$$c\boldsymbol{\sigma} \cdot \boldsymbol{p} \chi + (mc^2 + V)\varphi = E\varphi \tag{B.6a}$$

and

$$c\boldsymbol{\sigma} \cdot \boldsymbol{p}\varphi + (-mc^2 + V)\chi = E\chi. \tag{B.6b}$$

Because of equation (B.5), they can also be written as

$$-c\boldsymbol{\sigma} \cdot \boldsymbol{p}\chi + (mc^2 + V)\varphi = E\varphi \tag{B.7a}$$

and

$$-c\boldsymbol{\sigma} \cdot \boldsymbol{p}\varphi + (-mc^2 + V)\chi = E\chi. \tag{B.7b}$$

Suppose that the potential $V$ is replaced by its contrary number. Then, equations (B.6) become

$$c\boldsymbol{\sigma} \cdot \boldsymbol{p}\chi_{(-)} + (mc^2 - V)\varphi_{(-)} = E_{(-)}\varphi_{(-)} \tag{B.8a}$$

and

$$c\boldsymbol{\sigma} \cdot \boldsymbol{p}\varphi_{(-)} + (-mc^2 - V)\chi_{(-)} = E_{(-)}\chi_{(-)}. \tag{B.8b}$$

We multiply a minus sign to equation (B.8), and make exchanges $\chi_{(-)} \Leftrightarrow \varphi_{(-)}$ and $E_{(-)} \Leftrightarrow -E$, then the resultant equations are exactly the same as equations (B.7). It is concluded that in Dirac equation when the potential $V$ is replaced by $-V$, then the larger and smaller components exchanges and the obtained eigen values have a minus sign.

Actually, equation (B.4) has contained the NKE solutions for the contrary potential already. When multiplied by a minus sign, equation (B.4) immediately becomes

$$(-c\boldsymbol{\alpha} \cdot \boldsymbol{p} - mc^2\beta - V)\Psi = -E\Psi. \tag{B.9}$$

In this equation, the term $-mc^2\beta$ means that the two components $\varphi$ and $\chi$ exchanges. Therefore, it is seen that Dirac equation itself has contained the NKE solutions corresponding to contrary potential. The replacement of equation (B.5a) by equation (B.5b), actually means to exchange PKE and NKE solutions.

In appendix C, we will give the case of Coulomb potential as a specific example.

In conclusion, for any RQME, when the potentials takes the contrary number, the resultant eigenvalues do as well.

## Appendix C. The solution of a NKE particle subject to a Coulomb repulsive potential by Dirac equation

In this appendix, it is shown that an electron following Dirac equation in a repulsive Coulomb potential is of negative kinetic energy.

As a matter of fact, the formulas have been clearly given in textbooks, but NKE has never been paid enough attention. Here for the sake of simplicity, we just review the formulas in [40]. If there is only a scalar potential $V$ with spherical symmetry but no vector potential, the Hamiltonian of Dirac equation becomes

$$H = c\alpha_r P_r + \mathrm{i}\frac{\alpha_r}{r}c\beta\hbar k + mc^2\beta + V, \tag{C.1}$$

where

$$P_r = \frac{\boldsymbol{r} \cdot \boldsymbol{P} - \mathrm{i}\hbar}{r}, \quad \alpha_r = \frac{\boldsymbol{\alpha} \cdot \boldsymbol{r}\hbar}{r}. \tag{C.2}$$

Let the wave function be

$$\psi = \frac{1}{r}\begin{pmatrix} F(r) \\ G(r) \end{pmatrix}. \tag{C.3}$$

For the sake of convenience, the following denotations are defined.

$$\alpha_1 = \frac{mc^2 + E}{c\hbar}, \quad \alpha_2 = \frac{mc^2 - E}{c\hbar}, \tag{C.4}$$

$$\alpha = \sqrt{\alpha_1\alpha_2} = \frac{(m^2c^4 - E^2)^{1/2}}{c\hbar} \tag{C.5}$$

and

$$\rho = \alpha r. \tag{C.6}$$

The eigen equations become

$$-\left(\frac{\alpha_2}{\alpha} + \frac{V}{c\hbar\alpha}\right)F(\rho) + \left(\frac{d}{d\rho} + \frac{k}{\rho}\right)G(\rho) = 0 \tag{C.7a}$$

and

$$-\left(\frac{\alpha_1}{\alpha} - \frac{V}{c\hbar\alpha}\right)G(\rho) + \left(\frac{d}{d\rho} - \frac{k}{\rho}\right)F(\rho) = 0. \tag{C.7b}$$

These are equation (53.17) in [40]. Suppose that the electron is in a Coulomb attractive potential.

$$V(r) = -\frac{e^2}{r} = -\frac{\gamma\hbar c}{r}, \tag{C.8}$$

where $\gamma$ has been defined. With the potential (C.8), equation (C.7) become

$$-\left(\frac{\alpha_2}{\alpha} - \frac{\gamma}{\rho}\right)F(\rho) + \left(\frac{d}{d\rho} + \frac{k}{\rho}\right)G(\rho) = 0 \tag{C.9a}$$

and

$$-\left(\frac{\alpha_1}{\alpha} + \frac{\gamma}{\rho}\right)G(\rho) + \left(\frac{d}{d\rho} - \frac{k}{\rho}\right)F(\rho) = 0. \tag{C.9b}$$

The solved eigen energy is

$$E = \frac{mc^2}{\sqrt{1 + \gamma^2(\sqrt{k^2 - \gamma^2} + n')^{-2}}}. \tag{C.10}$$

Equation (C.10) is just equation (53.26) in [40], where $n'$ can be natural numbers. This energy includes three parts: static energy, kinetic energy and potential energy. The former two are positive and the last one is negative.

Now suppose that the electron is replaced by a positron. Then the positron is subject to a Coulomb repulsive potential

$$V(r) = \frac{e^2}{r} = \frac{\gamma\hbar c}{r} \tag{C.11}$$

instead of equation (C.8). Substituting equations (C.11) into (C.7), or simply taking replacement $\gamma \to \gamma$ in equation (C.9), we obtain

$$-\left(\frac{\alpha_2}{\alpha} + \frac{\gamma}{\rho}\right)F(\rho) + \left(\frac{d}{d\rho} + \frac{k}{\rho}\right)G(\rho) = 0 \tag{C.12a}$$

and

$$-\left(\frac{\alpha_1}{\alpha} - \frac{\gamma}{\rho}\right)G(\rho) + \left(\frac{d}{d\rho} - \frac{k}{\rho}\right)F(\rho) = 0. \tag{C.12b}$$

We take the following exchanges in equations (C.12):

$$k \to -k, \; E \to -E, \; \alpha_1 \Leftrightarrow \alpha_2 \text{ and } F(\rho) \Leftrightarrow G(\rho). \tag{C.13}$$

Then equation (C.12) goes back to the form of equations (C.9). Therefore, for the Coulomb repulsive potential equation (C.11), the two components of the eigen functions are exchanged, the total angular momentum is reversed and eigen energy becomes minus one, i. e.,

$$E = -\frac{mc^2}{\sqrt{1 + \gamma^2(\sqrt{k^2 - \gamma^2} + n')^{-2}}}. \tag{C.14}$$

The energy is contrary to that of (C.10). Its three parts, static energy, kinetic energy and potential energy, are all contrary to those of (C.10). Therefore, (C.14) is a NKE solution. The exchange $F(\rho) \Leftrightarrow G(\rho)$ apparently shows that the larger and smaller components exchange. For the Coulomb attractive (repulsive) potential, the PKE (NKE) particle has $F(\rho)$ ($G(\rho)$) as a larger component.

As a matter of fact, we have pointed out in appendix B that Dirac equation for stationary motion itself has already contained NKE solutions corresponding to the contrary potential. Specifically in the present case, equation (C.9) have contained the NKE solutions corresponding to the repulsive potential equation (C.11). This is easily shown by noticing that equation (C.9) remain invariant under the exchanges $\gamma \to -\gamma$ and equations (C.13). We emphasize that the NKE solution is a particle with a negative kinetic energy, but not an antiparticle with a positive kinetic energy.

Greiner [41] noticed that the spectrum (C.14) corresponded to the repulsive potential (C.11) and discussed the spectrum. Nevertheless, he did not realize that the kinetic energy of this solution was necessarily negative.

## Appendix D. Some topics of the NKE-related dark matter theory

In this appendix, we first shortly review the prevailing dark matter (DM) theory nowadays, and then present our viewpoints with regard to dark matter theory yielded from the concept of NKE.

The first signs of DM came in 1933, when Swiss astronomer Fritz Zwicky found that the galaxies in the Coma Cluster zinged through space so fast that their own gravity should not hold them together [42]. The history of finding DM was reviewed by Bertone and Hooper [43]. Researchers tried to detect DM by employing various means [44–54]. Experimental results were analyzed to give annual modulation [55]. Theoretical models of DM were postulated [56–60]. Possible properties that DM might have were guessed [60–69]. For example, there might be some kind of forces that DM acted [58]. A kind of DM called axion was assumed [70–75], and the attempts of detection of it were made [76]. Experimental measurements were suggested [77–80]. At least in some energy ranges, the exclusion of some of the DM that researchers had envisaged, including the axion [81–83], were determined. Nevertheless, effort has been still devoted to related theories [84] and experiment [85]. After the axion, the version of DM called weakly interacting massive particles (WIMPs) were proposed as the leading class of candidates of DM [86]. This assumption was to be confirmed [87], and the attempts of search of DM are still in undertaking [88, 89]. There have been other DM candidates suggested [90–93], but no one has been confirmed definitely. Up to now, what is DM on earth has not been known yet [93, 94].

The above theories assume that dark matter is some kinds of particles which have almost no interaction with the particles people have already known except gravitation. In the author's opinion, dark matter is all the particles we are dealing with now. They are dark simply when they are of NKE. When they are of PKE, they can be easily detected by various means people have developed. The NKE comes from Dirac equation as shown by equations (2.6)–(2.8). Therefore, our theory of dark matter originates from relativistic quantum equation. Besides, we do not make any postulation.

The details of NKE theory of dark matter are to be displayed. In the following, we briefly mention some, but not all, work in undertaking or to be done for NKE systems. They are listed point by point. Among them, some are problems within quantum mechanics itself.

1. From equations (2.20) and (2.21), the cases of a particle's energy less and greater than potential should be treated separately. Let us consider the harmonic oscillator potential $V(x) = \frac{1}{2}m\omega^2 x^2$. No matter how the energy is large, there are always regions where the potential is greater than the energy. This problem will be dealt with carefully.

2. After Dirac equation had been established, Klein evaluated the reflection coefficient of a Dirac particle encountering a one-dimensional step potential and found that the coefficient could be greater than 1, which was famous Klein's paradox. This paradox will be solved by making use of NKE solutions of Dirac equation and the evaluated reflection coefficient will never be greater than 1.

3. The scattering equation of a single NKE particle will be given. It will be in fact the mimic of that of a PKE particle that has been well established. It will involve the one-particle retarded and advanced Green's functions of a NKE free particle. For the time being, one key point can be pointed out: for a particle with PKE $K_{(+)} = p^2/2m$, there are two simple poles long the real axis in the complex $p$-plane, while for one with NKE $K_{(-)} = -p^2/2m$, there are two simple poles on the imaginary axis in the $p$-plane. Thus, there should be totally four poles in the complex $p$-plane. When a PKE particle is scattered, a distant observer sees that the scattering wave is a spherical wave in the form of

$$f(\theta, \varphi) \frac{e^{ipr}}{r}. \tag{D.1}$$

By contrast, the scattering wave function of a NKE particle will be in the form of

$$g(\theta, \varphi) \frac{e^{-kr}}{r}. \tag{D.2}$$

This wave function decay drastically with distance, and cannot reach instruments set in experiments. This may explain why up to now, in the field of elementary particle physics, experimental results have been in agreement with theories almost perfectly: the NKE particles have not been taken into account in the theories and could not be detected by scattering experiments.

Here, the scattering wave functions (D.1) and (D.2) are for low momentum motion. In elementary particle physics such as quantum electrodynamics, relativistic equations have to be utilized. Dirac equation has four independent solutions. Consequently, it is expected that there should be four poles in the $p$-plane, two belonging to PKE energy branch and other two to the NKE branch, coincident with the low momentum motion. The scattering equation of NKE wave functions of Dirac equation will be given.

    4. In section 3, it is seen that a NKE system may have discrete energy levels. Then, how do particles distribute in the levels? We will answer this question. The formalism of the statistical mechanics of systems composed of NKE particles will be presented. NKE Bosons and Fermions will be covered. It will be demonstrated that the concept of Dirac's Fermion Sea can be totally abandoned. The corresponding formalism of thermodynamics will also be given.

    5. One question should be answered that why the NKE systems are dark. When we say a matter is dark we mean that it is hardly to be detected by us. In point 3 above, it was mentioned that the scattering wave of a NKE particle exponentially decayed so that it was hardly detected. In point 4, it was mentioned that a NKE system might have energy levels. Subsequently, transitions between the NKE energy levels would naturally occur. This kind of transition events can neither be detected by the present apparatus, and we will tell why, see point 13 below.

    6. In our observable universe, matters can be macroscopic and microscopic, obeying formalism of classical mechanics and quantum mechanics, respectively. It is believed that dark matter can also be macroscopic and microscopic. The latter obey the formalism of quantum mechanics we have already know. The formalism of the macroscopic dark bodies will be derived. Actually, they also follow Newton's three laws of mechanics.

    7. A gas composed of NKE molecules without interactions between them except collisions is called a dark ideal gas. The molecular kinetics of a dark ideal gas will be given. From both the molecular kinetics and statistics mechanics mentioned in point 4, a NKE system produces negative pressure. Astrophysicists know that our universe inflates with an acceleration now, and a negative pressure should be responsible for this. Nevertheless, they have not known where the negative pressure was from. Our NKE theory can tell that one source of negative pressure is NKE systems (the other source is from dark energy).

    8. The collision between PKE and NKE particles will be investigated. This is a way the author can suppose to probe the NKE particles for the time being. It was mentioned in point 5 that neither the scattering wave of a NKE particle nor the transitions between NKE energy levels could be detected. The collision between PKE and NKE particles may be an effective means to detect NKE particles. In section 4 of the present paper, we have suggested experiments to let photons collide NKE electrons. Later, other collisions will be proposed.

    9. In sections 3 and 4, it was mentioned that two NKE electrons can constitute a stable system by means of their Coulomb repulsive interaction. More few-body systems will be studied, which are composed of few NKE particles, e. g., that containing one NKE proton and two electrons with the same kind of charges. The systems constituted by one PKE and one NKE particles will also be researched.

    10. Many-body theory for NKE systems will be developed. The formalism will be almost parallel to that for PKE systems. As an example, let us consider a PKE electron gas in a solid state material. Because both the electrons' kinetic energy and the Coulomb interaction energy are positive, a negative energy is needed for such an electron system to be stable. This is the Coulomb attraction between the electrons and ions in the solid. By contrast, for NKE electrons, their kinetic energy is negative and the Coulomb interacting energy are positive. The positive and negative energies may make the NKE electron system stable, without need a third energy to balance the system.

    11. In quantum electrodynamics, the expression of probability density of spin-0 particle, equation (2.22) here, is incorrect, and the possible poles in the *p*-plane arising from NKE energy branch have not been taken into account. Therefore, it is probable that some formulas in quantum electrodynamics need modified. The possible influence on quantum field theory should also be analyzed.

    12. It is well known that in weak interaction, parity is not conserved. It actually means that for PKE particles people can probe, the parity is not conserved. There may be such a possibility that NKE particles also participate the reaction events, and the total parity is conserved. The NKE particles may take some parities away. This is to be investigated.

    13. At last, we mention our theory with respect to dark energy which is in developing. The most of the space in our universe looks dark. It is believed that in the dark space, there are dark matters, i. e., NKE matters. The speeds of stars, gravitational lenses and other observable evidences have revealed the existence of dark matters. There should be a variety of systems composed of NKE particles and they ought to have various activities such as transitions between energy levels by absorbing and releasing energies. If they release energies, we should have detected some, but we have not. We will give the reason why we have not been able to detect the energies the dark systems release. This will involve our theory of dark energy. By the way, according to our theory, dark energy also contributes negative pressure.

    Up to now, we have used energies to detect the activities of PKE particles. That is to say, PKE particles match the energy. By contrast, NKE particles match dark energy. Our universe is really symmetric with regard to the observed and dark ingredients!

## ORCID iDs

Huai-Yu Wang  https://orcid.org/0000-0001-9107-6120